\documentclass{aastex63}
\usepackage{makecell}
\usepackage{amsmath}
\usepackage{cases}
\graphicspath{{./}{figures/}}
\begin{document}
%\begin{CJK}{UTF8}{}

%\title{Detection of the extended $\gamma$-ray emission around Calvera}
\title{Detection of the extended $\gamma$-ray emission from the new supernova remnant G321.3-3.9 with Fermi-LAT}

\author[0009-0004-5450-8237]{Xiaolei Guo}
\author{Xi Liu}
\affiliation{School of Physical Science and Technology, Southwest Jiaotong University, Chengdu 610031, China\\
\href{mailto:xlguo@swjtu.edu.cn}{xlguo@swjtu.edu.cn}}

%\affil
%\email{ylxin@swjtu.edu.cn (YLX)}



\begin{abstract}

With the 15 yrs of Pass 8 data recorded by the {\em Fermi} Large Area Telescope, we report the detection of an extended gigaelectronvolt emission component with a 68\% containment radius of $0^{\circ}\!.85$, which is
spatially associated with the newly identified supernova remnant (SNR) G321.3-3.9.
The $\gamma$-ray spectrum is best described by a log-parabola model in the energy range of 100 MeV - 1 TeV, which shows a significant spectral curvature at $\sim$ 1 GeV. 
Either a leptonic or a hadronic model could explain the multi-wavelength data of G321.3-3.9, while the leptonic model predicts a too low strength of magnetic field.
Also considering the flat radio spectrum of G321.3-3.9 and the $\gamma$-ray upper limit in the low energy band, the hadronic model is favored.
The spatial coincidence between the $\gamma$-ray morphology and the diffuse thermal X-ray emission of G321.3-3.9 and the curved gigaelectronvolt $\gamma$-ray spectrum of it make G321.3-3.9 to be similar to the typical middle-aged SNRs interacting with molecular clouds.
Such characteristics provide another evidence of the potential hadronic origin for its $\gamma$-ray emission.
While there is no molecular cloud detected around G321.3-3.9, which challenges the hadronic model.

\end{abstract}

%% to the KEY in the \bibitem in the reference list below. ,，

\section{Introduction}
\label{intro}

It is widely believed that supernova remnants (SNRs) are the
main accelerators of Galactic cosmic rays (CRs) with energies
up to the ``knee'' feature of the CR spectrum \citep{1934PNAS...20..259B, 2022RvMPP...6...19L}.
Such scenario is supported by the non-thermal X-ray
emission detected in many SNRs, which indicates the
acceleration of electrons to hundreds of TeV energies \citep[e.g.,][]{1995Natur.378..255K,2007Natur.449..576U}.
In addition, the Large Area Telescope (LAT) on board the {\em Fermi} satellite has detected the characteristic pion-decay signature from several middle-aged $\gamma$-ray SNRs interacting with
molecular clouds \citep[MCs;][]{2013Sci...339..807A,2016ApJ...816..100J}, which is considered to be the most direct
evidence of nuclei acceleration in SNRs.
So far, nearly three hundred of SNRs have been clearly identified mainly through the radio observations \footnote{http://snrcat.physics.umanitoba.ca/SNRtable.php} \citep{2014BASI...42...47G}, 
while only ten percent among them is detected in $\gamma$-rays \citep[e.g.,][]{2019ApJ...874...50Z, 2021ApJ...910...78Z,2022ApJ...941..194X,2023ApJ...955...84X,2023ApJ...951..142Z}.
Searching for the $\gamma$-ray emission from the individual remnant would be helpful to study the acceleration of particles in SNRs, and further to evaluate contributions of SNRs to the flux of Galactic CRs.
Here we report on the $\gamma$-ray detection of the newly identified SNR G321.3-3.9.

%G321.3-3.9 was firstly classified as a SNR candidate using a polarimetric radio continuum survey at 2.4 GHz taken with the Parkes 64-m radio telescope \citep{1997MNRAS.287..722D}. It has a partial elliptical structure with a well-defined arc on the southeastern side. The radio size of G321.3-3.9 was determined to be about 1.8$^\circ$ $\times$ 1.3$^\circ$. Afterwards, with the 843 MHz observation from the Molonglo Galactic Plane Survey 2nd Epoch (MGPS-2), \citet{2014PASA...31...42G} also classified G321.3-3.9 to be a SNR candidate, and the radio morphology of it is described as an elliptical and almost complete shell with size of 1.8$^\circ$ $\times$ 1.1$^\circ$.
\citet{1997MNRAS.287..722D} classified G321.3-3.9 as an SNR candidate using Parkes data at 2.4 GHz. 
Using data from the second epoch of the Molonglo Galactic Plane Survey (MGPS-2) at 843 MHz, \citet{2014PASA...31...42G} confirmed this result and improved its morphology to an oval, almost full shell shape of 1.8$^\circ$ $\times$ 1.1$^\circ$ size.
Recently, \citet{2024arXiv240117294M} identified G321.3-3.9 as a new SNR with the data collected by several radio surveys, including the Galactic Plane Monitoring
(GPM) campaign across 185-215 MHz conducted with the Murchison Widefield Array \citep[MWA;][]{2013PASA...30....7T,2018PASA...35...33W}, MGPS-2 at 843 MHz \citep{1981PASA....4..156M}, the Continuum HI Parkes All-Sky Survey (CHIPASS) at 1.4 GHz \citep{2014PASA...31....7C}, and the S-band Polarisation All Sky Survey (S-PASS) at 2.3 GHz \citep{2019MNRAS.489.2330C}.
These data span a frequency range from 200 MHz to
2300 MHz, and quantify the structure of G321.3-3.9 as an elliptical shape with major and minor axes of about 1.7$^\circ$ $\times$ 1.1$^\circ$.
With the measured flux density from CHIPASS (11.5$\pm$0.6 Jy at 1.4 GHz) and S-PASS (7.6$\pm$0.5 Jy at 2.3 GHz), \citet{2024arXiv240117294M} calculated a radio spectral index of $\alpha \simeq$ $-$0.8 $\pm$ 0.4, which is consistent with synchrotron emission from an expanding shell in the Sedov-Taylor phase of SNRs.
With the X-ray data from eROSITA, an X-ray diffuse structure is detected filling almost the entire radio shell, and the spectral analysis models the X-ray emission to be totally thermal.
By comparing the X-ray and optical extinction maps of G321.3-3.9, the distance of it is suggested to be about 1 kpc, which is also consistent with the absorption column obtained from the X-ray spectral analysis of the remnant \citep{2024arXiv240117294M}.
Based on the distance of 1 kpc and a low ISM density of 0.1 cm$^{\rm -3}$, the age of G321.3-3.9 is estimated to be about 4 kyr using the SNR evolutionary model in \citet{2017AJ....153..239L}.
It should be noted that the age calculation of SNR could be effected by many factors, the most important of which being the distance uncertainty and the local medium density.

In this work, we carry out a detailed analysis of the gigaelectronvolt $\gamma$-ray emission in the direction of the SNR G321.3-3.9 with the {\em Fermi}-LAT Pass 8 data.
In Section 2, the data analysis routines and results are presented, including the spatial and spectral analyses. 
A discussion of the origin of the non-thermal emission based on the multi-wavelength data is presented in Section 3, followed by the conclusions in Section 4.

\section{Fermi-LAT Data Analysis}
\label{fermi}

\subsection{Data Reduction}
\label{data reduction}

We select the latest Pass 8 version of the {\em Fermi}-LAT data with ``Source'' event class, recorded from 2008 August 4 (Mission Elapsed Time 239557418) to 2023 August 4 (Mission Elapsed Time 712800005).
The region of interest (ROI) we analyzed is a $20^\circ \times 20^\circ$ square region centered at the position of the SNR G321.3-3.9 (R.A. = $233^{\circ}\!.121$, Dec = $-60^{\circ}\!.892$).
For the spectral analysis, the energy range is adopted to be 100 MeV - 1 TeV, 
while the spatial analysis is carried out in the energy range of 1 GeV - 1 TeV considering the improved point-spread function (PSF) at higher energies.
The events whose zenith angles are larger than $90^\circ$ are excluded to reduce the contamination from Earth Limb.
The data are analyzed with the standard {\it ScienceTools} software package \footnote {http://fermi.gsfc.nasa.gov/ssc/data/analysis/software/} with the binned analysis method, together with the instrumental response function (IRF) of ``P8R3{\_}SOURCE{\_}V3''.
The Galactic and isotropic diffuse background models used here are {\tt gll\_iem\_v07.fits} and {\tt iso\_P8R3\_SOURCE\_V3\_v1.txt}
\footnote {http://fermi.gsfc.nasa.gov/ssc/data/access/lat/BackgroundModels.html}, respectively.
The sources in the incremental version of the fourth {\em Fermi}-LAT source catalog \citep[4FGL-DR4;][]{2022ApJS..260...53A,2023arXiv230712546B}, together with the two diffuse backgrounds, are included in the model.
During the fitting procedure, the spectral parameters and the normalizations of sources within $5^{\circ}$ around the SNR G321.3-3.9 are set to be free, together with the normalizations of the two diffuse backgrounds.

\subsection{Spatial Analysis}
\label{Spatial}

\begin{figure*}[!htb]
	\centering
	\includegraphics[width=0.48\textwidth]{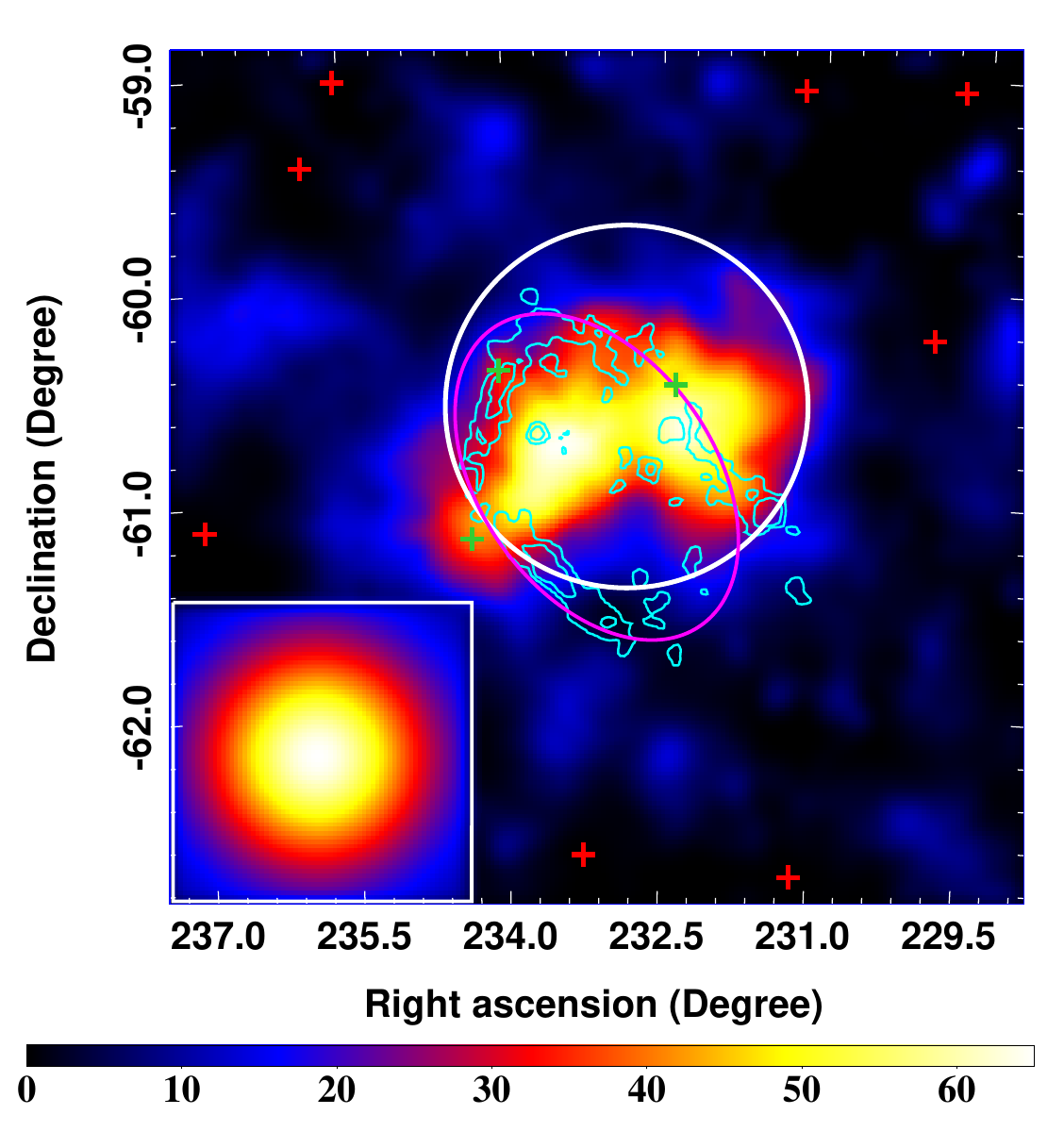}
	\includegraphics[width=0.48\textwidth]{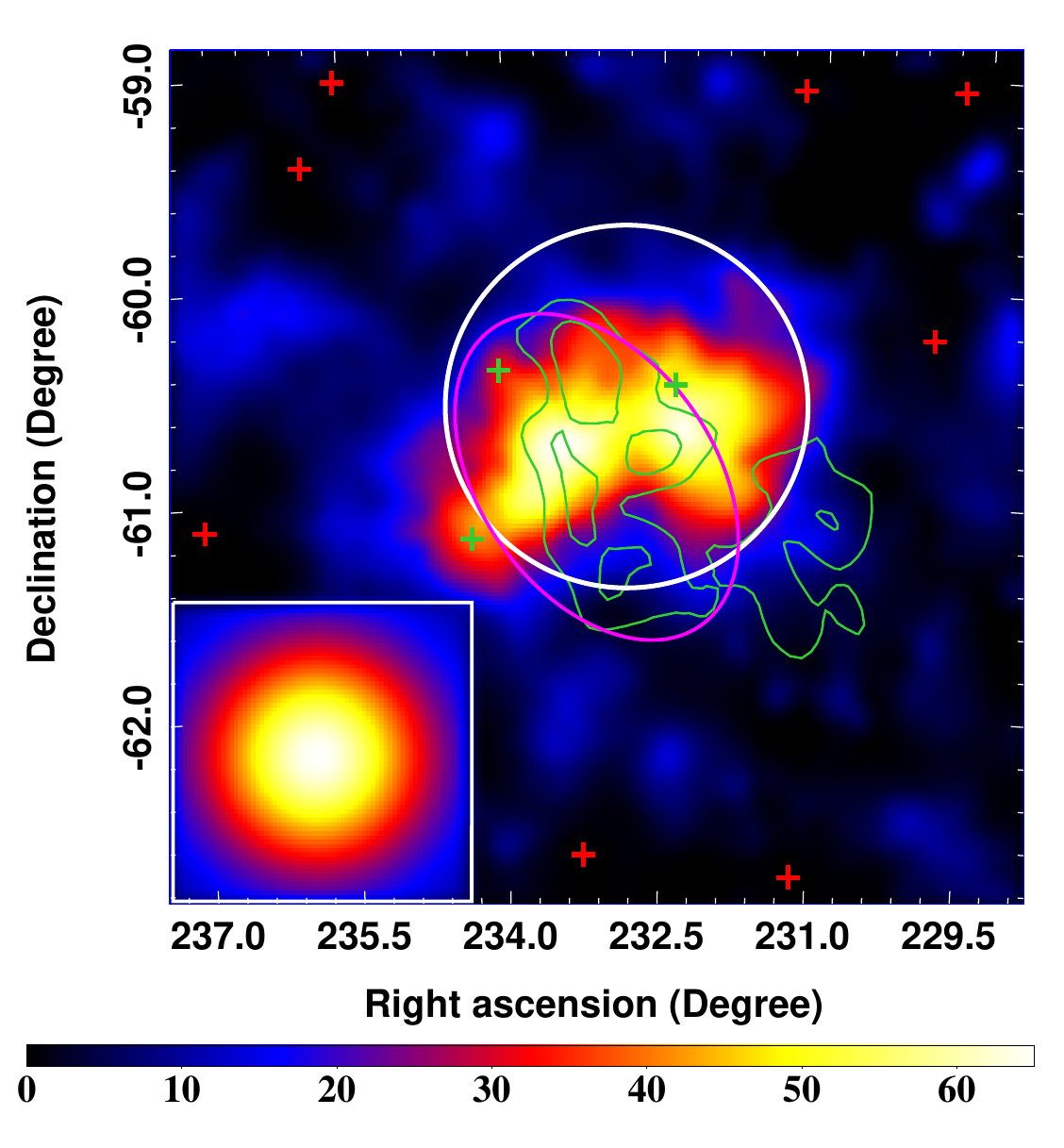}
	\caption{$4^{\circ}\!.0$ $\times$ $4^{\circ}\!.0$ TS map for photons above 1 GeV smoothed with a Gaussian kernel of $\sigma$ = $0^{\circ}\!.04$.	The red crosses show the background and/or foreground sources in the 4FGL-DR4 catalog and the green crosses mark the three 4FGL-DR4 point sources in the region of G321.3-3.9. The white solid circle shows the extension of R$_{\rm 68}$ for the 2D-Gaussian template, and the magenta ellipse shows the radio size of G321.3-3.9 \citep{2024arXiv240117294M}. The cyan contours describe the elliptical-shaped radio shell of the SNR G321.3-3.9 by Parkes-MIT-NRAO (PMN) radio survey at 4.85 GHz \citep{1993AJ....105.1666G}, and the green contours show the X-ray image in the eROSITA all-sky surveys from 0.2 to 1.0 keV \citep{2024arXiv240117294M}. The inset of the panels in the white square shows the simulated LAT PSF in the same energy range, adopting the same smoothing.} %The yellow circle represents the 68\% containment size of the PSF at the energy threshold of each map.}
	\label{fig1:tsmap}
\end{figure*}

In the 4FGL-DR4 catalog, there are three $\gamma$-ray point sources (4FGL J1536.2-6023, 4FGL J1529.4-6027, 4FGL J1537.3-6110) in the region of G321.3-3.9, and all of them have no identified counterparts \citep{2020ApJS..247...33A, 2022ApJS..260...53A}. 
With the data from 1 GeV to 1 TeV, we first performed the maximum likelihood analysis and created a $4^{\circ}\!.0$ $\times$ $4^{\circ}\!.0$ Test Statistic (TS) map with the command {\em gttsmap} by subtracting the modelled emission from the diffuse backgrounds and all 4FGL-DR3 sources (except the three 4FGL sources above) in the best-fit model, which is shown in Figure \ref{fig1:tsmap}. 

The TS map shows significant $\gamma$-ray excess at and around G321.3-3.9.
To study the excess $\gamma$-ray emission, we test several
different spatial models in the energy range of 1 GeV - 1 TeV, which are summarized in Table \ref{table:spatial}.
%Taking into account the two $\gamma$-ray spots shown in Figure \ref{fig1:tsmap}, 
We first considered two point sources model to describe the $\gamma$-ray emission in this region, and the initial positions of the two sources are adopted to that of pixels with the maximum TS value in Figure \ref{table:spatial}.
The best-fit coordinates of the two point sources are optimized by {\tt Fermipy}, a {\tt PYTHON} package that automates analyses with {\it ScienceTools} \citep{2017ICRC...35..824W}, which is performed by maximizing the model likelihood with	respect to source position.
Then a model with three point sources is also tested following the 4FGL-DR4 catalog.
In addition, we also adopted the extended templates to describe the $\gamma$-ray emission around G321.3-3.9, including a uniform disk and a two-dimensional (2D) Gaussian template.
The best-fit central coordinates and extensions marked by the radius containing 68\% of the intensity (R$_{\rm 68}$) of them are derived using {\tt Fermipy}, which are listed in Table \ref{table:spatial}.
%For the template of the uniform ellipse, the major and minor axes are adopted to be $1^{\circ}\!.7$ and $1^{\circ}\!.1$, together with the inclination angle of $125^{\circ}$, which follows the radio observation by GPM at $\sim$200 MHz \citep{2024arXiv240117294M}.
The radio image in Parkes-MIT-NRAO (PMN) radio survey at 4.85 GHz \citep{1993AJ....105.1666G} and X-ray image in eROSITA all-sky surveys from 0.2 to 1.0 keV \citep{2024arXiv240117294M} are also adopted to be the spatial templates.
The values of the overall maximum likelihood for the different spatial templates are listed in Table \ref{table:spatial}.
We adopt the Akaike information criterion \citep[AIC;][]{1974ITAC...19..716A}\footnote{AIC = -2ln$\mathcal{L}$+2{\em k}, where $\mathcal{L}$ is the value of overall maximum likelihood and {\em k} is the numbers of the degree of freedom of model.} to compare the goodness of the fit in the different models by calculating $\Delta$AIC, the difference between the two point sources model and others.
It is evident from Table \ref{table:spatial} that the 2D-Gaussian model provides the minimum $\Delta$AIC value, which is adopted as the spatial template in the following spectral analysis.
The central position and the 68\% containment radius of the 2D-Gaussian template are fitted to be R.A. = $232^{\circ}\!.836$, Dec = $-60^{\circ}\!.562$ and R$_{\rm 68}$ = $0^{\circ}\!.850$, respectively.
With the extended template of 2D-Gaussian model, the TS value is fitted to be about 157.9 in the energy range of 1 GeV - 1 TeV, corresponding to a significance level of $\sim$11.8 $\sigma$ with five degrees of freedom.
It should be noted that the spatial template with X-ray image gives the nearly identical results with that of the 2D-Gaussian model.

\begin{table}[!htb]
\centering
\caption {Spatial Analysis for the $\gamma$-ray emission around G321.3-3.9 between 1 GeV and 1 TeV}
\begin{tabular}{ccccccc}
\hline \hline
Spatial Template    &  R.A., Dec       & Degrees of Freedom   & -log(Likelihood)  & $\Delta$AIC\\
%                    & & R$_{\rm 68}$
% &                               &$\rm 10^{-11}$ ph/$\rm cm^{2}$/s  &  Value &  Freedom   \\
\hline
2 Point Sources   & \makecell[c]{$233^{\circ}\!.806 \pm 0^{\circ}\!.040$, $-61^{\circ}\!.003 \pm 0^{\circ}\!.054$ \\ $232^{\circ}\!.153 \pm 0^{\circ}\!.065$, $-60^{\circ}\!.657 \pm 0^{\circ}\!.084$}   & 8  & -743577.68 & 0\\
\hline
3 Point Sources   & \makecell[c]{$233^{\circ}\!.547 \pm 0^{\circ}\!.038$, $-60^{\circ}\!.838 \pm 0^{\circ}\!.038$ \\ $232^{\circ}\!.150 \pm 0^{\circ}\!.053$, $-60^{\circ}\!.652 \pm 0^{\circ}\!.085$\\ $234^{\circ}\!.267 \pm 0^{\circ}\!.080$, $-61^{\circ}\!.176 \pm 0^{\circ}\!.053$}   & 12  & -743585.47 & -7.58\\
\hline
Uniform disk  & \makecell[c]{$232^{\circ}\!.763 \pm 0^{\circ}\!.077$, $-60^{\circ}\!.717 \pm 0^{\circ}\!.106$ \\ R$_{\rm 68}$ = $0^{\circ}\!.804 ^{+0^{\circ}\!.057}_{-0^{\circ}\!.061}$}   & 5  & -743598.80 & -48.24 \\
\hline
2D-Gaussian   & \makecell[c]{$232^{\circ}\!.836 \pm 0^{\circ}\!.095$, $-60^{\circ}\!.562 \pm 0^{\circ}\!.136$ \\ R$_{\rm 68}$ = $0^{\circ}\!.850 ^{+0^{\circ}\!.064}_{-0^{\circ}\!.065}$}   & 5  & -743601.42  & -53.48\\
\hline
%Uniform ellipse   & \makecell[c]{$233^{\circ}\!.121$, $-60^{\circ}\!.892$\\ major axis = $1^{\circ}\!.7$, minor axis = $1^{\circ}\!.1$\\ inclination angle =  $125^{\circ}$}  & 2  & -743579.47  & -15.58\\
%\hline
Radio image   & --  & 0  & -743588.65  & -37.94\\
\hline
X-ray image   & --  & 0  & -743595.64  & -51.92\\
\hline
\end{tabular}
\label{table:spatial}
\end{table}   

\subsection{Spectral Analysis}
\label{Spectral}

With the spatial template of the 2D-Gaussian model, we performed a spectral analysis of the $\gamma$-ray emission of G321.3-3.9 in the energy range of 100 MeV - 1 TeV.
We first fit the data using a single power-law (PL; $dN/dE \propto E^{-\alpha}$) spectrum, and the TS value is fitted to be $\rm TS_{PL}$ $\sim$ 133.3. 
%And the spectral index and integrated photon flux of SrcX are fitted to be 1.67 $\pm$ 0.10 and $(6.16 \pm 2.10)\times10^{-10}$ photon cm$^{-2}$ s$^{-1}$, respectively.
And
we also adopted a log-parabola (LogPb; $dN/dE \propto E^{-(\alpha+\beta {\rm log}(E/E_b))}$) spectrum to test the spectral
curvature of G321.3-3.9.
The TS value is fitted to be $\rm TS_{LogPb} \sim$ 186.8.
The higher TS value with a log-parabola model
shows a significant spectral curvature of the $\gamma$-ray emission from G321.3-3.9, and the variation of TS values, $\rm \Delta TS$ = $\rm TS_{LogPb}$ - $\rm TS_{PL}$ $\simeq$ 53.5, corresponds to $\sim$7.3$\sigma$
improvement with respect to a power-law model with one
additional degree of freedom.
The fitting of log-parabola yields the spectral parameters of $\alpha = 1.56 \pm 0.07$ and $\beta = 0.67 \pm 0.06$ at $\rm E_b$ = 1 GeV. And the integrated photon flux is calculated to be $(7.36 \pm 0.52)\times10^{-9}$ photon cm$^{-2}$ s$^{-1}$, respectively.

\begin{figure*}[!htb]
	\centering
	\includegraphics[width=0.6\textwidth]{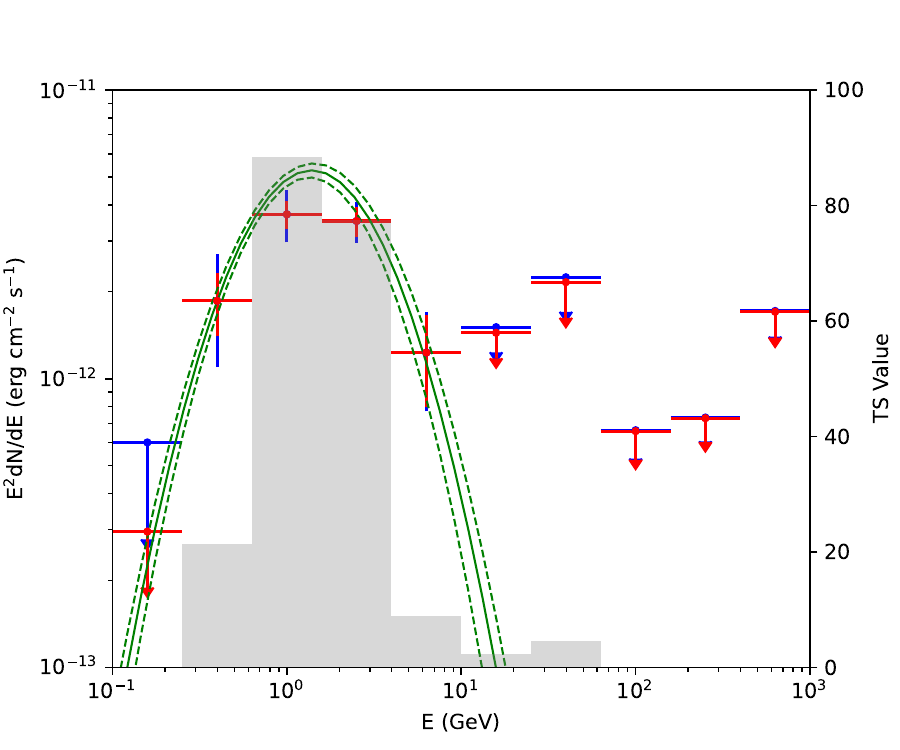}
	\caption{The $\gamma$-ray SED of G321.3-3.9 in the energy range of 100 MeV - 1 TeV. The red error bars demonstrate the statistical errors. The sums of the statistical and systematic errors calculated by $\sigma$ = $\sqrt{\sigma_{\rm stat}^2 + \sigma_{\rm syst}^2}$ are marked by the blue error bars. The flux data with 95\% upper limits are shown as the arrows, and the gray histogram indicates the TS value for each energy bin. The best-fit log-parabola spectral model in 100 MeV - 1 TeV is shown as the green solid line, together with its 1$\sigma$ statistic error shown as the green dashed lines.}
	\label{fig2:sed}
\end{figure*}

Furthermore, to drive the $\gamma$-ray spectral energy distribution (SED) of G321.3-3.9, we divide the data from 100 MeV to 1 TeV into ten logarithmically even energy bins, and perform the same likelihood fitting analysis for each energy bin.
For the energy bin with a TS value lower than 5.0, an upper limit with a 95\% confidence level is calculated.
Taking into account the effect of the Galactic diffuse emission, the systematic errors are also calculated for the SED by changing the best-fit normalization of the Galactic diffuse model artificially by $\pm$6\% \citep{2010ApJ...714..927A}.
The sums of the statistical and systematic errors are calculated by $\sigma$ = $\sqrt{\sigma_{\rm stat}^2 + \sigma_{\rm syst}^2}$ for each energy bin.
The results of the $\gamma$-ray SED are shown in Figure \ref{fig2:sed}, and the global fitting results with the log-parabola model from 100 MeV to 1 TeV are also plotted with the green solid and dashed lines.

\section{Discussion}
\label{discussion}

\begin{figure*}[!htb]
	\centering
	\includegraphics[width=0.6\textwidth]{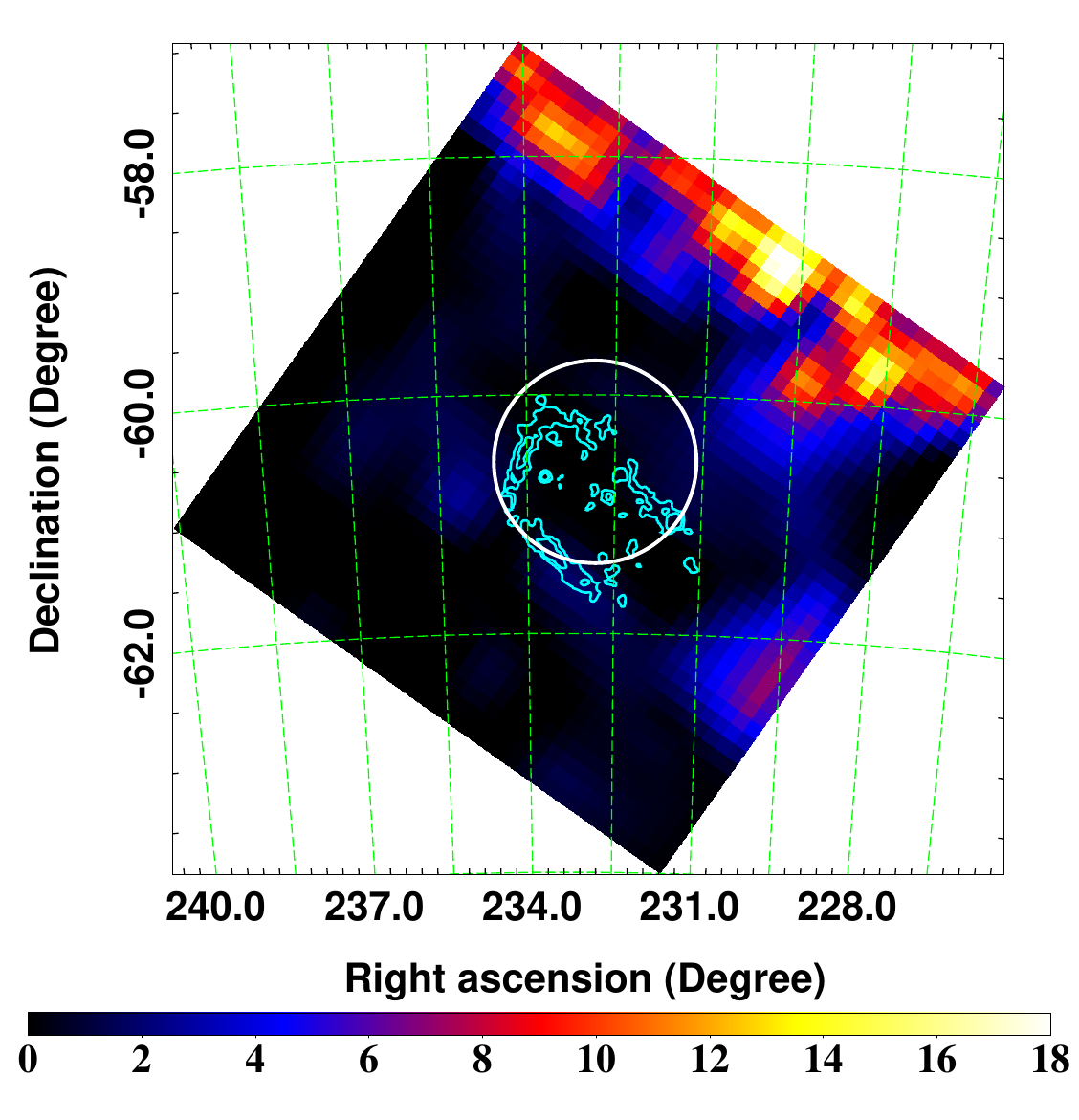}
	\caption{The velocity-integrated CO emission intensity with the unite of K km s$^{-1}$ around the SNR G321.3-3.9 \citep{2001ApJ...547..792D}. The 68\% containment radius of the extended $\gamma$-ray emission is marked by the white solid circle. The cyan contours show the radio image of G321.3-3.9 at 4.85 GHz, as shown in the left panel of Figure \ref{fig1:tsmap}.}
	\label{fig3:COmap}
\end{figure*}

The above spatial data analysis shows the extended $\gamma$-ray emission in the direction of the SNR G321.3-3.9, whose extension is well consistent with the radio size of the remnant.
The spatial coincidence suggests that the extended $\gamma$-ray source could be the GeV counterpart of the SNR G321.3-3.9, although such possibility could be existing that the $\gamma$-ray emission could be from the sea cosmic-rays interacting with high density gas that happens to be partially spatial coincident with the remnant.
Considering the results of spatial analysis, the $\gamma$-ray morphology is better coincident with the diffuse thermal X-ray emission compared with the radio shell morphology, which fills a large part of the inner remnant region and partly overlaps with the remnant’s radio emission in the west and southeast \citep{2024arXiv240117294M}.
Such spatial correlation among the $\gamma$-ray and the thermal X-ray
emission also exists in other SNRs, like W51C \citep{2009ApJ...706L...1A} and the Cygnus Loop \citep{2011ApJ...741...44K}.
The 2D-Gaussian template provides the best-fit result for the $\gamma$-ray emission of the SNR G321.3-3.9 with the 68\% containment radius of $0^{\circ}\!.850$.

For the ambient environment of the SNR G321.3-3.9, we try to search for the components of molecular cloud in a large extend using the data from the CfA 1.2m millimeter-wave telescope \citep{2001ApJ...547..792D}, which is shown in Figure \ref{fig3:COmap}.
While no bright CO emission is detected in the $\gamma$-ray-emitting region, which could be attributed to dissociation
of molecules by the radiative precursor of the SNR due to the
photoionization and photodissociation effects \citep{2008A&A...480..439P}.
Also we can't rule out the possibility of the existing of CO-dark gas in this region, which can not be traced by CO observations \citep{2005Sci...307.1292G}.

\begin{figure*}[!htb]
	\centering
	\includegraphics[width=0.6\textwidth]{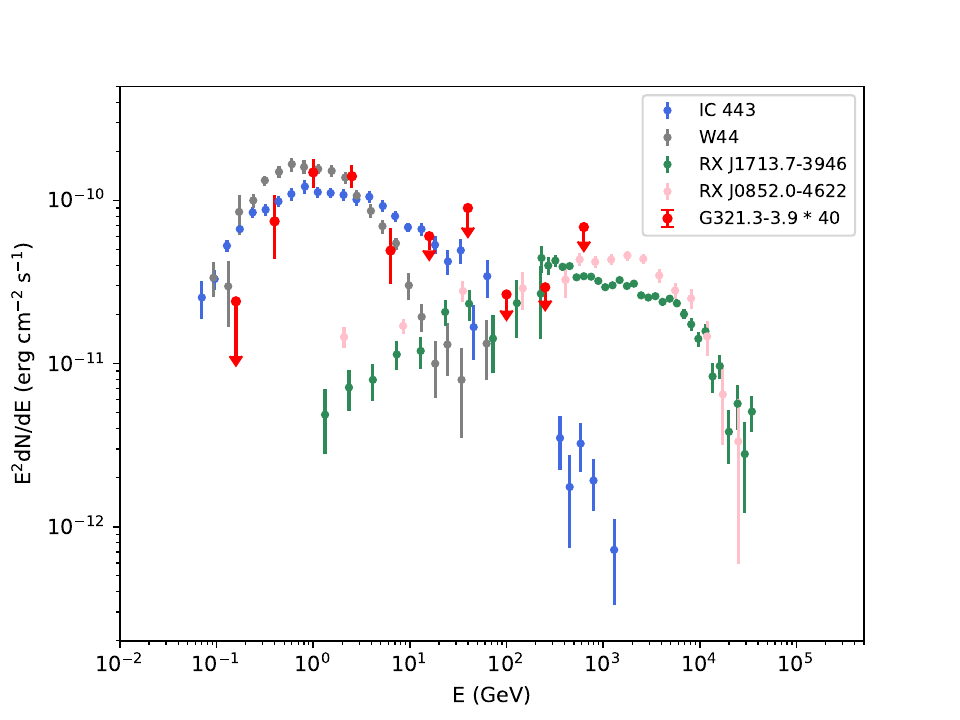}
	\caption{Typical $\gamma$-ray SEDs of various $\gamma$-ray luminous SNRs \citep{2013Sci...339..807A, 2018A&A...612A...6H, 2018A&A...612A...7H}. The energy fluxes of the SNR G321.3-3.9 are scaled upward by 40 times for spectral shape comparison with other $\gamma$-ray SNRs.}
	\label{fig4:sed-multiSNRs}
\end{figure*}

In order to compare the energetics of the SNR G321.3-3.9 with other typical $\gamma$-ray SNRs, we plot the $\gamma$-ray spectra of several SNRs detected in gamma-rays with {\em Fermi}-LAT, as shown in Figure \ref{fig4:sed-multiSNRs}.
The $\gamma$-ray SNRs could be roughly divided into two classes based on their $\gamma$-ray spectra. 
One class of SNRs is the young-aged system \citep[age $\sim$ 1000-2000 yrs;][]{2020pesr.book.....V} with bright non-thermal X-ray emission and hard gigaelectronvolt spectrum, like RX J1713.7-3946 \citep{2018A&A...612A...6H} and RX J0852.0-4622 \citep{2018A&A...612A...7H}.
The $\gamma$-ray emission from most of these SNRs is suggested to be produced by inverse Compton scattering (ICS) of accelerated electrons (leptonic model), and their $\gamma$-ray spectra shows the spectral curvature at TeV energies. 
Another class is the middle-aged SNR interacting with molecular clouds, including IC 443 and W44 \citep{2013Sci...339..807A}. 
The $\gamma$-ray spectra of these SNRs shows the spectral curvatures at $\sim$ GeV, and most of their $\gamma$-ray emission is suggested to be from the decay of neutral pions produced by the inelastic {\em pp} collisions (hadronic model).
The spectral analysis above shows a significant spectral curvature at $\sim$ GeV for the $\gamma$-ray spectrum of G321.3-3.9, which is similar to that of the typical middle-aged SNRs.
The $\gamma$-ray luminosity of G321.3-3.9 from 100 MeV - 1 TeV is calculated to be about 1.4 $\rm \times 10^{33}$ $\rm (d/1kpc)^{2}$ $\rm erg$ $\rm s^{-1}$.
Such value is about two magnitudes lower than that of IC 443 and W44, which makes the SNR G321.3-3.9 to be among the faintest SNRs identified
with {\em Fermi}-LAT \citep{2012APh....39...22T}.

\begin{figure*}[!htb]
	\centering
	\includegraphics[width=0.48\textwidth]{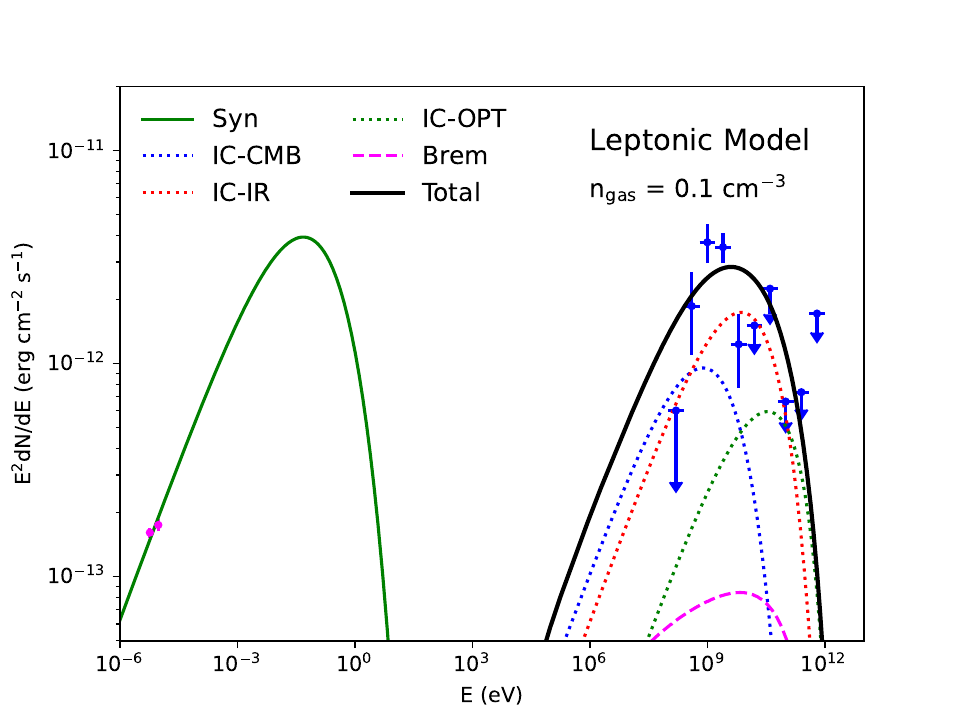}
	\includegraphics[width=0.48\textwidth]{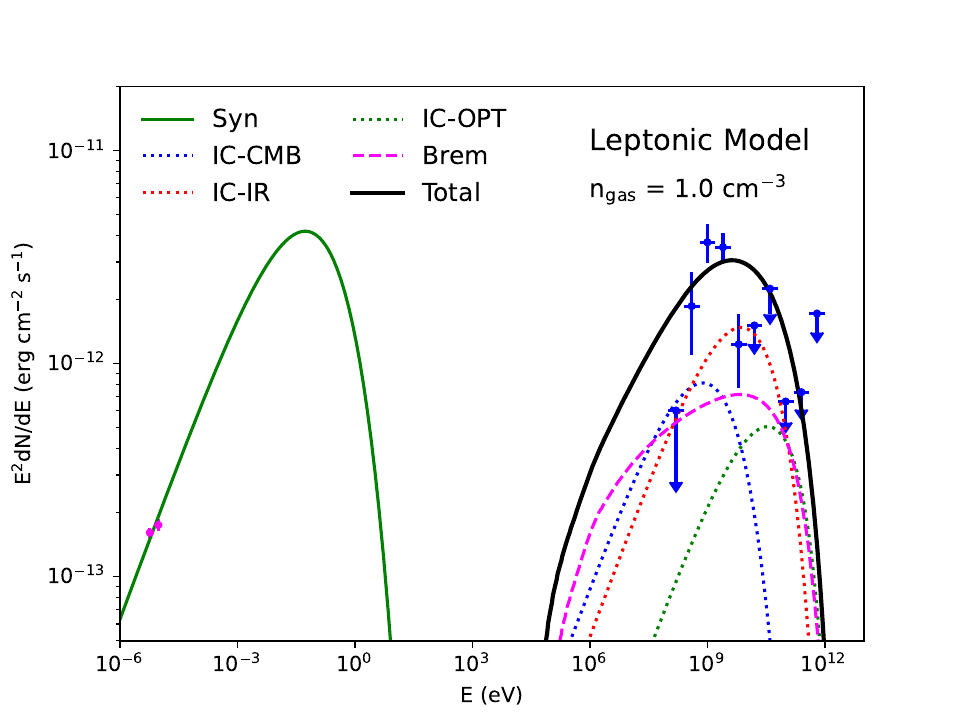}\\
	\includegraphics[width=0.48\textwidth]{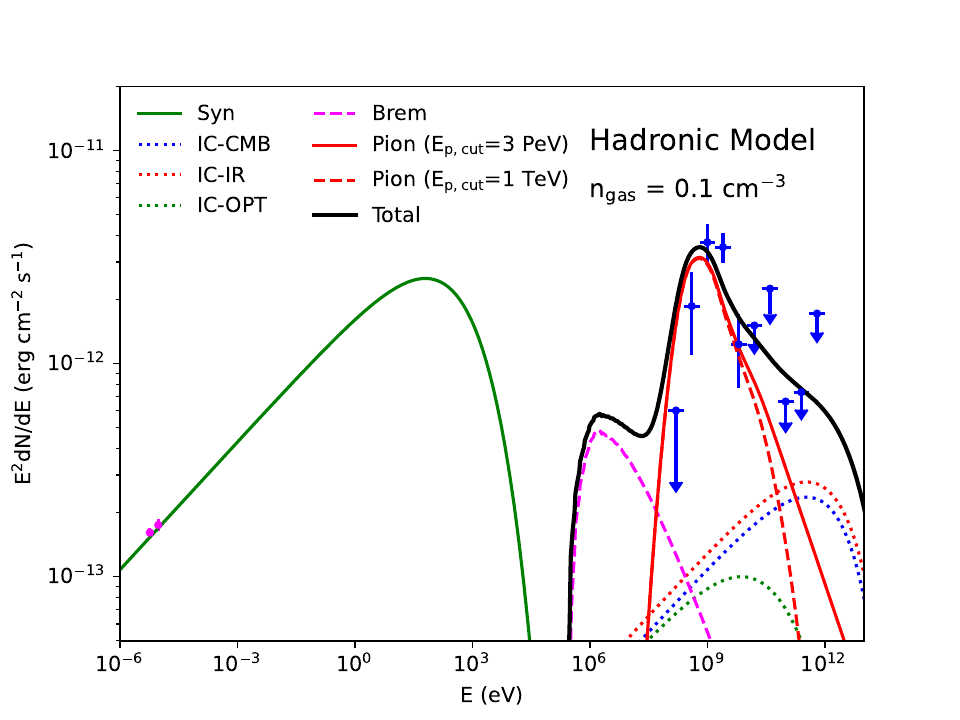}
	\includegraphics[width=0.48\textwidth]{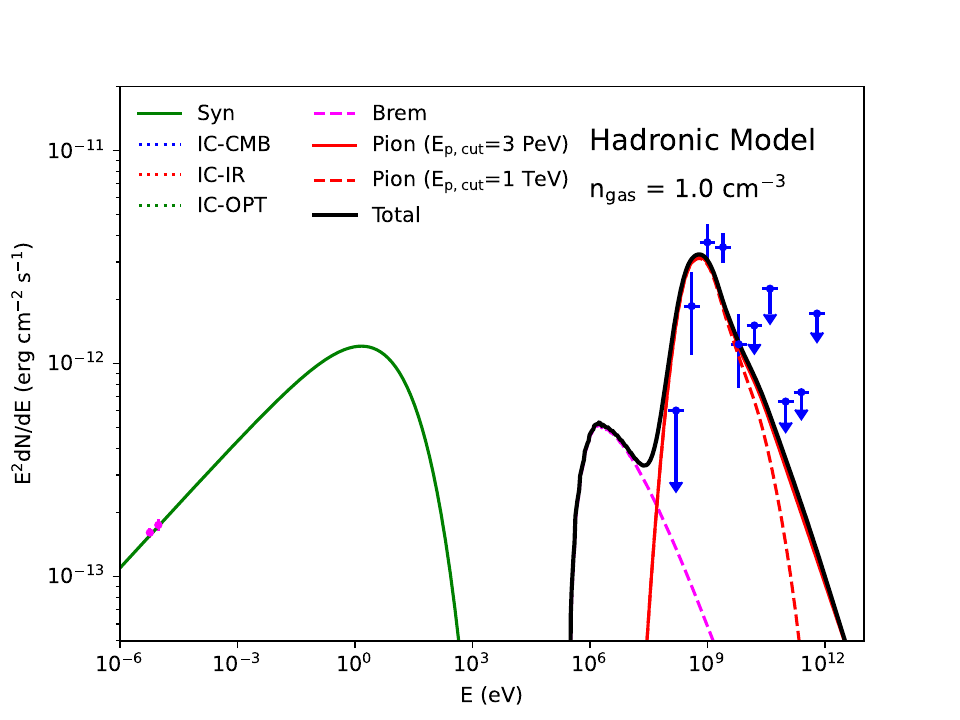}
	\caption{The multi-wavelength SED of the SNR G321.3-3.9. The top panels are for the leptonic models with the different values of gas density, and the bottom panels are for the hadronic models. The radio data shown as the magenta dots are from \citet{2024arXiv240117294M}. The green solid line shows the synchrotron component. The blue, red and green dotted lines represent ICS component of the different seed photons, shown as the legend. The bremsstrahlung component is shown as the magenta dashed line. The black solid curve represents the sum of the different radiation components for the $\gamma$-ray emission. In the bottom panels, the red solid and dashed lines indicate the hadronic models with $\rm E_{\rm p,cut}$ = 3 PeV and $\rm E_{\rm p,cut}$ = 1 TeV, and the contributions from ICS components are too low to show in the bottom right panel.}
	\label{fig5:model}
\end{figure*}

To reveal the origin of $\gamma$-ray emission from the SNR G321.3-3.9, we quantitatively discuss the leptonic and hadronic models here.
The spectrum of electrons and/or protons is assumed to be a power-law model with an exponential cutoff in the form of
$dN/dE_i \propto E_i^{-\alpha_i} \exp(-E_i/E_{i,\rm cut})$, 
where $\alpha_i$ and $E_{i, \rm cut}$ are the spectral index and the cutoff energy, respectively, for $i = e$ or $p$.
The radio data produced by high-energy electrons via the synchrotron process are adopted from the observations by CHIPASS and S-PASS in \citet{2024arXiv240117294M}.
Besides the cosmic microwave background (CMB), the infrared (IR) and optical (OPT) photon fields are also considered for the ICS process. 
And the temperatures and energy densities of them are calculated by the empirical relation introduced in \citet{2011ApJ...727...38S} to be about (T$_{\rm IR}$ = 30 K, u$_{\rm IR}$ = 0.5 eV cm$^{-3}$) and (T$_{\rm OPT}$ = 3000 K, u$_{\rm OPT}$ = 1.0 eV cm$^{-3}$), respectively. %\citep{2006ApJ...648L..29P, 2008ApJ...682..400P}.
The distance of the SNR G321.3-3.9 is adopted to be 1 kpc \citep{2024arXiv240117294M}.
The different values of gas density with $\rm n_{gas}$ = 0.1 cm$^{-3}$ and 1.0 cm$^{-3}$ are used to calculate the contributions from bremsstrahlung and hadronic processes. 
The components of the different radiation mechanisms are computed using {\em naima} package \citep{2015ICRC...34..922Z}, and the fitting results with leptonic and hadronic models are compiled in Figure \ref{fig5:model}. 

For the leptonic model, the radio spectral index of $-$0.8 $\pm$ 0.4 corresponds to a spectral index of electrons to be $\sim$ 2.6 for the synchrotron process and a $\gamma$-ray spectral index of $\sim$ 1.8 for ICS component.
However, the low flux values in the first two bins of the $\gamma$-ray spectrum suggests a harder distribution for electrons.
The spectral index of electrons is adopted to be 2.0 considering the statistic uncertainties of the radio data.
The $\gamma$-ray spectrum and the upper limits in the high energy band constrain the total energy of electrons above 1 GeV to be W$_{\rm e}$ $\sim$ (3.0 - 3.5) $\times$ 10$^{\rm 47}$ erg and a cutoff energy of E$_{\rm e, cut}$ $\sim$ 0.4 TeV.
To explain the radio flux, the magnetic field strength of $\sim$ (7.0 - 7.5) $\mu$G is needed.
Such value is lower than that of the typical shell type SNRs with tens of $\mu$G \citep{2019ApJ...874...50Z,2021ApJ...910...78Z}.
The value of magnetic field strength and the cutoff energy suggest the synchrotron cooling timescale of electrons to be about 6 $\times$ 10$^{\rm 5}$ yrs, which is much larger than the age of the SNR G321.3-3.9 with 4 kyr \citep{2024arXiv240117294M}.
As shown in the top panels of Figure \ref{fig5:model}, the leptonic model would produce a much higher contribution at the energy of $\sim$ 100 MeV compared to the detected $\gamma$-ray upper limit.

In the hadronic model, the electronic spectral index is set to be 2.6, which is consistent with the radio spectral index of -0.8 for the SNR G321.3-3.9.
To reduce the number of free parameters, the spectral index of protons is also set to be 2.6, and the cutoff energy of electrons is calculated by equalling the synchrotron cooling timescale and the age of the SNR G321.3-3.9 with 4 kyr, which is given by
E$_{\rm e, cut}$ = 1.25 $\times$ 10$^{\rm 7}$ t$_{\rm age; yr}^{-1}$ B$_{\rm \mu G}^{\rm -2}$ TeV \citep{2008ApJ...676.1210Z}. 
As shown in the bottom panels of Figure \ref{fig5:model}, the contribution of bremsstrahlung component and the $\gamma$-ray upper limit in the low energy band constrain the maximum value of the total energy of electrons to be about W$_{\rm e}$ $\sim$ 1.4 $\times$ 10$^{\rm 47}$ erg and $\sim$ 1.4 $\times$ 10$^{\rm 46}$ erg for the gas density of 0.1 cm$^{-3}$ and 1.0 cm$^{-3}$, respectively.
The corresponding magnetic field strengths are about $\sim$ 10$\mu$G and $\sim$ 35$\mu$G to explain the radio flux, which could be regarded as the lower limits of magnetic field strength.
The cutoff energies of electrons are calculated to be E$_{\rm e, cut}$ $\sim$ 31 TeV and 2.5 TeV, respectively.
The cutoff energy of protons cannot be well constrained and is first adopted to be the energy of the cosmic ray knee with $\rm E_{\rm p,cut}$ = 3 PeV \citep{2022RvMPP...6...19L}.
Also considering the fact that the cut-off energy of protons in the middle-aged SNRs is usually much lower than PeV \citep{2023A&A...679A..22S}, we adopt a low cutoff energy of protons of $\rm E_{\rm p,cut}$ = 1 TeV to estimate the different hadronic models. 
The total energy of protons is calculated to be $\sim (1.4-1.5) \times 10^{50}(n_{\rm gas}/0.1\ {\rm cm}^{-3})^{-1}$ erg, which suggests an acceleration efficiency of $\sim$ 15\% compared to the explosion energy of a typical supernova ($\sim$ $10^{51}$ erg). 

Compared with the leptonic model, the hadronic model predicts a lower flux in the energy of $\sim$ 100 MeV, which is more consistent with the non-detection of the $\gamma$-ray emission around 100 MeV \citep{2022PhRvL.129g1101F}.
Also the hadronic model predicts a higher magnetic field strength than the leptonic model.
And also considering the spatial coincidence between the $\gamma$-ray emission and the diffuse thermal X-ray emission, the hadronic model is favored for G321.3-3.9 \citep{2008JCAP...01..018K}, which is similar to SNR Puppis A \citep{2017ApJ...843...90X}.
However, the absence of molecular cloud around G321.3-3.9 challenges the hadronic model, and the further studies of the ambient environment of this remnant is necessary.

\begin{table*}
	\centering
	%\tabletypesize{\small}
	\caption {Parameters for the models}
	\begin{tabular}{cccccccccc}
		\hline \hline
		Model & n$_{\rm gas}$ & $\alpha_e$ & $\alpha_p$  & E$_{e,\rm cut}$ & E$_{p, \rm cut}$ & W$_e$  & W$_p$ & $B$ \\
		& (cm$^{\rm -3}$)	&  &   & (TeV)          & (TeV)        & ($10^{47}$ erg) & ($10^{50}$ erg) & ($\mu$G) \\
		\hline
		Leptonic  &$0.1$ & $2.0$  & $-$  & $0.4$ & $-$ & $3.5$ & $-$ & $7.0$ \\
		&$1.0$ & $2.0$  & $-$  & $0.4$ & $-$ & $3.0$ & $-$ & $7.5$ \\
		\hline
		Hadronic  &$0.1$ & $2.6$ & $2.6$ & $31.3$  & $1/3000$  & $1.4$  & $1.4/1.5$ & $10$ \\
		&$1.0$ & $2.6$ & $2.6$ & $2.5$   & $1/3000$  & $0.15$ & $0.14/0.15$ & $35$ \\
		\hline
		\hline
	\end{tabular}
	\label{table:model}
	\tablecomments{The total energy of relativistic particles, $W_{e,p}$, is calculated for $E > 1$ GeV. The two values of $W_{p}$ correspond to the different values of E$_{p, \rm cut}$ with 1 TeV or 3 PeV.}
\end{table*}

\section{Conclusions}
\label{con}

Using 15 yrs of Pass 8 data recorded by the {\em Fermi}-LAT, we report the detection of the extended $\gamma$-ray emission around the SNR G321.3-3.9.
The $\gamma$-ray morphology can be well described by a 2D-Gaussian template with the 68\% containment radius of $0^{\circ}\!.85$, which is well consistent with the radio size of the remnant.
The fitting results of the spatial analysis suggest that the $\gamma$-ray emission of G321.3-3.9 is spatially well coincidence with the diffuse thermal X-ray emission, filling a large part of the inner remnant region and partly overlapping with the remnant’s radio emission in the west and southeast.
The $\gamma$-ray spectrum of G321.3-3.9 follows a log-parabola model in the energy range of 100 MeV - 1 TeV, which shows a significant spectral curvature at $\sim$ 1 GeV with $\sim$7.3$\sigma$ improvement with respect to a power-law model.
%The $\gamma$-ray morphology and the curved spectrum make G321.3-3.9 to be similar to the typical middle-aged SNRs interacting with molecular clouds, although there no significant CO emission detected around the remnant.
%whose $\gamma$-ray emission are suggested to be from the hadronic process.
Both leptonic and hadronic models are considered to explain the multi-wavelength data of G321.3-3.9, and neither of them can be strictly ruled out.
The leptonic model predicts a too low strength of magnetic field compared with that of the typical shell type SNRs, and the flat radio spectrum seems to be not well consistent with the $\gamma$-ray upper limit in the low energy band for the leptonic model.
Such results above suggest that the hadronic model is favored.
The $\gamma$-ray morphology and the curved spectrum of G321.3-3.9 make it to be similar to the typical middle-aged SNRs interacting with molecular clouds,
which provides another evidence of the potential hadronic origin for its $\gamma$-ray emission.
However, the absence of molecular cloud around G321.3-3.9 challenges the hadronic model, and the further multi-wavelength observations for G321.3-3.9, especially the molecular cloud searching around it, would be helpful to test the hadronic model and clarify the origin of its $\gamma$-ray emission.

\acknowledgments

We would like to thank the anonymous referee for very helpful comments, which help to improve the paper. 
We thank
Silvia Mantovanini and Werner Becker for providing the information about the X-ray image of the SNR G321.3-3.9. 
This work is based on observations made with NASA's Fermi Gamma-Ray Space Telescope, 
and supported by the Natural Science Foundation of Sichuan Province of China under grant Nos.
2024NSFSC0452, the Fundamental Research Funds for the Central Universities
under grant Nos. 2682024CG002,
and the National Natural Science Foundation of China under grant Nos. 12103040.




\bibliography{sample63}{}
\bibliographystyle{aasjournal}


%\end{CJK}
\end{document}